\documentclass[aps,prl,twocolumn,superscriptaddress,floats,a4paper,nobibnotes]{revtex4}
\usepackage{newfloat}
\usepackage{latexsym}
\usepackage{dcolumn}
\usepackage{graphicx}
\usepackage{amssymb}
\usepackage{amsmath}
\usepackage{float}
\usepackage{hyperref}
\hypersetup{colorlinks,linkcolor=blue,citecolor=blue,urlcolor=blue}
\usepackage[left=18mm,right=18mm,top=25mm,bottom=25mm]{geometry}
\usepackage[version=4]{mhchem}
\usepackage{bm}
\usepackage{color}
\usepackage[english]{babel}

\begin{document}

\title{Microscopic theory for electron hydrodynamics in monolayer and bilayer graphene}

\author{Derek Y. H. Ho}
\affiliation{Department of Physics and Centre for Advanced 2D Materials, National University of Singapore, 2 Science Drive 3, 117551, Singapore}

\author{Indra Yudhistira}
\affiliation{Department of Physics and Centre for Advanced 2D Materials, National University of Singapore, 2 Science Drive 3, 117551, Singapore}

\author{Nilotpal Chakraborty}
\affiliation{Yale-NUS College, 16 College Avenue West, 138614, Singapore}

\author{Shaffique Adam}
\email{shaffique.adam@yale-nus.edu.sg}
\affiliation{Department of Physics and Centre for Advanced 2D Materials, National University of Singapore, 2 Science Drive 3, 117551, Singapore}
\affiliation{Yale-NUS College, 16 College Avenue West, 138614, Singapore}

\date{\today}

\begin{abstract}
Electrons behave like a classical fluid with a momentum distribution function that varies slowly in space and time when the quantum mechanical carrier-carrier scattering dominates over all other scattering processes.  Recent experiments in monolayer and bilayer graphene have reported signatures of such hydrodynamic electron behavior in ultra-clean devices.  In this theoretical work, starting from a microscopic treatment of electron-electron, electron-phonon and electron-impurity interactions within the Random Phase Approximation, we demonstrate that monolayer and bilayer graphene both host two different hydrodynamic regimes.  We predict that the hydrodynamic window in bilayer graphene is stronger than in monolayer graphene, and has a characteristic `v-shape' as opposed to a `lung-shape'.  Finally, we collapse experimental data onto a universal disorder-limited theory, thereby proving that the observed violation of Wiedemann-Franz law in monolayers occurs in a regime dominated by impurity-induced electron-hole puddles.
\end{abstract}
\maketitle

For most metals, electron-like quasiparticles with Fermi-Dirac statistics move ballistically like classical billiard balls between collision centers, while the scattering crosssection during the collisions are treated quantum mechanically. These properties are captured by solving the Boltzmann equation for the semiclassical distribution function. In the unusual case where the carrier-carrier scattering dominates over all other kinds of scattering, a coarse-graining procedure transforms the Boltzmann equation into hydrodynamic quantities governed by classical fluid mechanics~\cite{landau_course_1981}. Various proposals for effects that are peculiar for electrons can manifest themselves in such a hydrodynamic regime including a non-monotonic temperature dependence of resistivity that marks the transition from boundary-limited to viscous flow~\cite{gurzhi_hydrodynamic_1968}, an electron Bernoulli effect~\cite{govorov_hydrodynamic_2004}, spontaneous excitation of plasma oscillations~\cite{dyakonov_shallow_1993}, signatures of electron viscosity in magnetotransport~\cite{avron_viscosity_1995,alekseev_negative_2016}, and faster than ballistic electron transport~\cite{guo_higher-than-ballistic_2017}.

In spite of more than 50 years of theorizing about electron hydrodynamics, experimental observation has been sparse and disputed (see e.g.~Ref.~\cite{de_jong_hydrodynamic_1995}).  The reason is that in many experimental configurations electron-impurity scattering dominates at low temperatures while electron-phonon scattering dominates at high temperatures providing a small window in temperature (if any) for observing such electron hydrodynamics.  The experimental situation changed recently with reports of charge carriers behaving hydrodynamically in monolayer~\cite{bandurin_negative_2016, crossno_observation_2016} and bilayer graphene~\cite{bandurin_negative_2016,nam_electronhole_2017} (see also Ref.~\cite{moll_evidence_2016} that reported a large viscous contribution to the resistance of palladium cobaltate). 

\begin{figure*}[!th]
\begin{center}$
\begin{array}{cc}
\includegraphics[height=!,width=6.5cm] {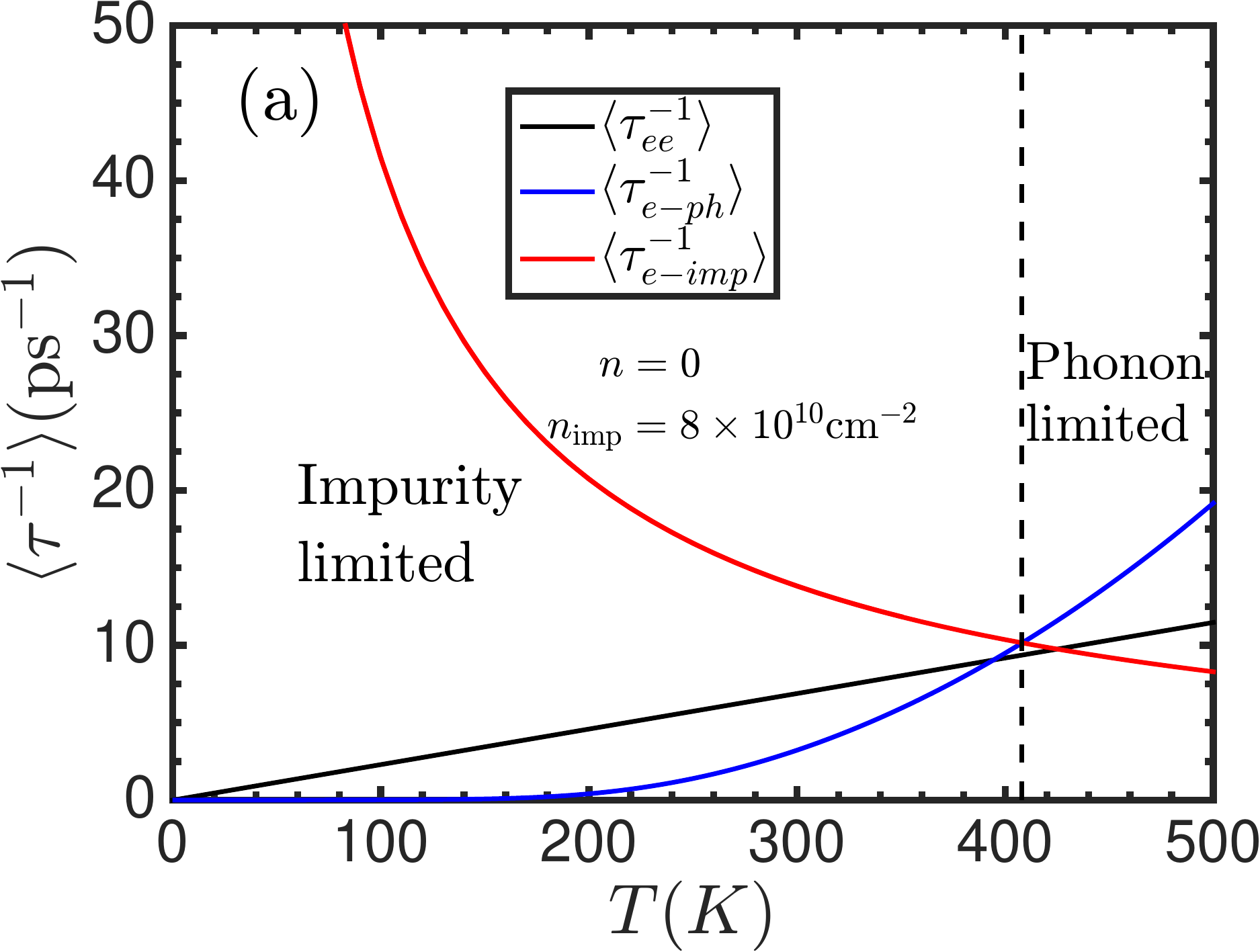} &
\includegraphics[height=!,width=6.5cm] {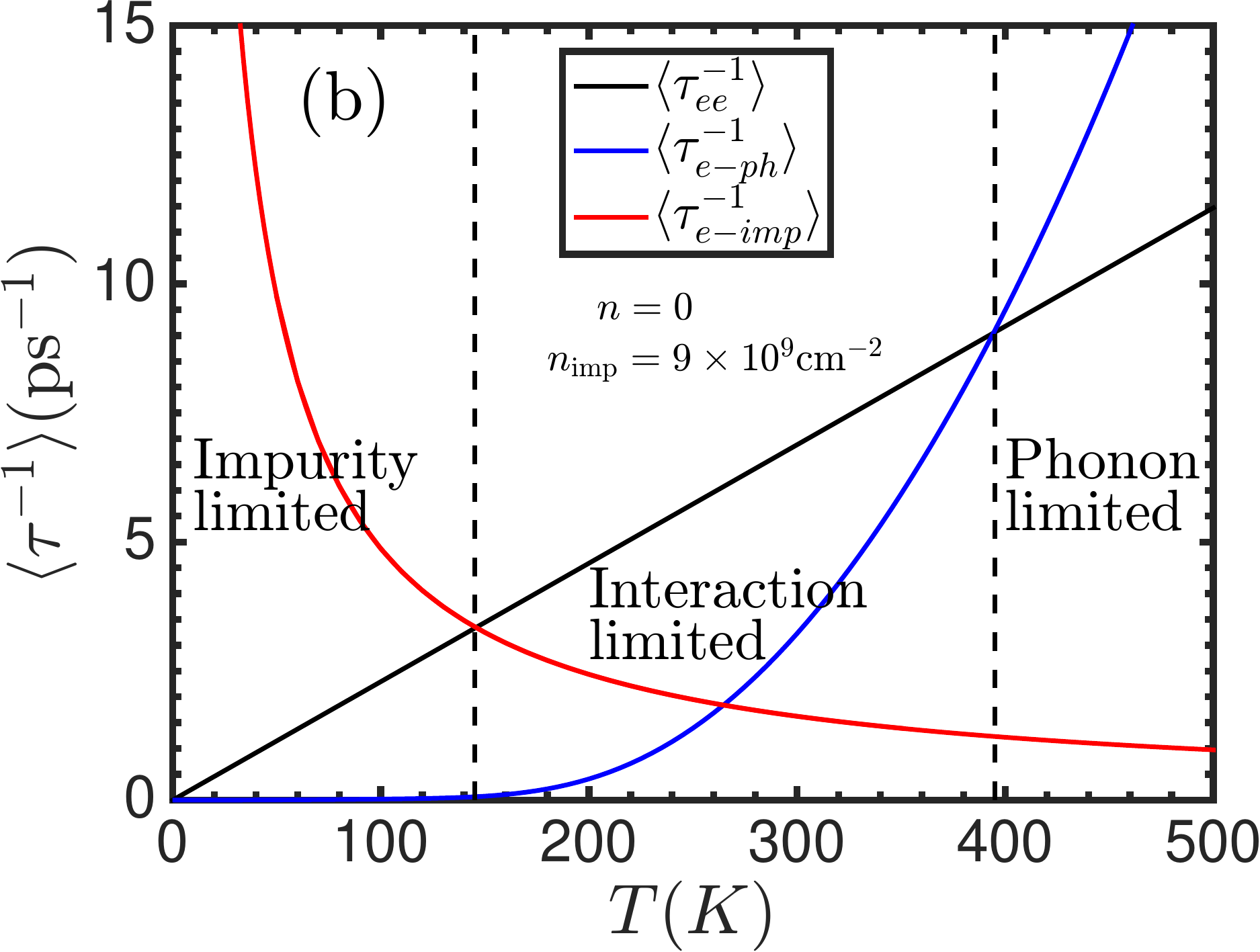} 
\end{array}$
\end{center}
\caption{(Color online) Opening of the hydrodynamic window in charge neutral graphene.  (a) For disordered graphene, or graphene on a silica substrate, impurity scattering dominates at low temperature, while phonon scattering dominates at high temperature and there is no window where the quasiparticle lifetime is limited by carrier scattering.  (b) For ultraclean graphene on h-BN, typical impuritiy concentrations are $10^{10}~{\rm cm}^{-2}$.  Impurity scattering still dominates at low temperature and phonons at high temperature, but a robust window opens up at intermediate temperatures where electron-electron scattering rate dominates.  In this hydrodynamic window, electronic properties are described by fluid mechanics.} \label{Fig1}
\end{figure*}

It is important to note that gapless (or narrow band-gap) materials support two very different types of hydrodynamic regimes depending on whether temperature is larger or smaller than the Fermi energy.  At low temperature, only one carrier contributes to the hydrodynamic Fermi liquid and the transport can be described using a single macroscopic charge current evolving under the standard Navier-Stokes equation~\cite{torre_nonlocal_2015}.  This is in contrast to high temperature, where both electrons and holes contribute to a hydrodynamic plasma 
that involves three coupled macroscopic currents~\cite{narozhny_hydrodynamics_2015,lucas_transport_2016}.  It is curious that while the Manchester and Harvard groups used samples of comparable quality, the former observed only single fluid hydrodynamics, and the latter only the coupled plasma regime.

In this Letter, we show that for monolayer and bilayer graphene, a large and robust hydrodynamic window exists over a wide-range of carrier densities and temperatures supporting both the plasma and single-carrier hydrodynamic regimes (for monolayer, this requires impurity concentrations lower than $\sim 10^{11}~{\rm cm}^{-2}$, while no such constraint applies for bilayers).  We predict very different hydrodynamic windows for monolayer and bilayer graphene revealing the subtle competition between different scattering mechanisms.  The difference between the experiments is that the negative non-local resistance measurement at Manchester (which is a proxy for viscous backflow) is observable only in the single-carrier regime, while the violation of Wiedmann-Franz law seen at Harvard occurs in the electron-hole puddle regime dominated largely by charged-impurity scattering.  Our calculation of the interaction-limited quasiparticle lifetime in bilayer graphene shows a universal scaling collapse in agreement with the observation in Geneva of electron-hole scattering limited transport~\cite{nam_electronhole_2017}.  Taken together, our work provides the first microscopic theory for the hydrodynamic regime in graphene monolayers and bilayers. 

Only a careful comparison of the various scattering mechanisms can decide when an electronic system is described by the hydrodynamic equations.  
We therefore look at the contribution to the quasiparticle lifetime $\tau(\varepsilon) = \hbar / 2 \mathrm{Im} \left[\Sigma(\varepsilon)\right]$, where $\Sigma(\varepsilon)$ is the corresponding self-energy calculated within the Random Phase Approximation (RPA) for that scattering process.  At a given carrier density $n$ and temperature $T$, the contribution of a given scattering mechanism to the quasiparticle lifetime is obtained by thermal-averaging~\cite{schutt_coulomb_2011} 
\begin{equation}
\langle \tau^{-1} \rangle = \frac{\int^{\infty}_{-\infty} d\varepsilon~\varepsilon~\frac{\partial n_{F}(\varepsilon-\mu)}{\partial \varepsilon}~\tau^{-1}(\varepsilon)}{\int^{\infty}_{-\infty} d\varepsilon~\varepsilon~\frac{\partial n_{F}(\varepsilon-\mu)}{\partial \varepsilon} } \label{th-avrg},
\end{equation}
where $n_{F}$ is the Fermi-Dirac distribution.  The chemical potential $\mu$ is found by solving $n = n_{e} - n_{h}$~\cite{adam_temperature_2010}.
For monolayer graphene, the self-energy $\Sigma^{\rm RPA}(\varepsilon)$ for various scattering mechanisms is known in the literature (see e.g.~Ref.~\cite{polini_quasiparticle_2014} for electron-electron (ee) scattering, Ref.~\cite{hwang_single-particle_2008} for electron-impurity (e-imp) scattering and Ref.~\cite{sohier_phonon-limited_2014} for electron-phonon (e-ph) scattering), however to the best of our knowledge, we are the first to do the thermal averaging for the e-ph and e-imp contributions.  For bilayer (discussed in more detail below), we calculate for the first time the  self-energy for the e-e scattering mechanism within the RPA, in addition to performing the thermal average for e-ph and e-imp contributions.  In the plasma regime where there are both electron and hole carriers, we count all appropriate channels i.e.~$\langle \tau^{-1}_{cc}\rangle$ = $\langle \tau^{-1}_{ee}\rangle$ + $\langle \tau^{-1}_{eh}\rangle$ + $\langle \tau^{-1}_{hh}\rangle$ (which for notational simplicity, we generically call $\langle \tau^{-1}_{ee} \rangle$ for the remainder of this Letter).  
The above procedure determines $\langle \tau^{-1}_{ee} \rangle$, $\langle \tau^{-1}_{\mathrm{e-imp}} \rangle $ and $\langle \tau^{-1}_{\mathrm{e-ph}} \rangle$.  We then identify the collision-dominated regime as when $\langle \tau^{-1}_{ee} \rangle > \langle \tau^{-1}_{\mathrm{e-imp}} \rangle, \langle \tau^{-1}_{\mathrm{e-ph}} \rangle$.

Figure \ref{Fig1} illustrates our results for the case of monolayer graphene at the Dirac point. Fig.~\ref{Fig1}(a) is the typical case for graphene on silicon dioxide, where impurity scattering dominates at low temperature and the $A_1'$ phonon mode dominates at high temperature. There is no window for which $\langle \tau^{-1}_{ee} \rangle$ dominates.  This explains why there have been no reports of graphene hydrodynamics until very recently.  Fig.~\ref{Fig1}(b) shows typical impurity concentrations of ultraclean graphene on h-BN.  In this case, a hydrodynamic window opens up at intermediate temperatures.     

\begin{figure*}
\begin{center}$
\begin{array}{cc}
\includegraphics[height=!,width=7.5cm] {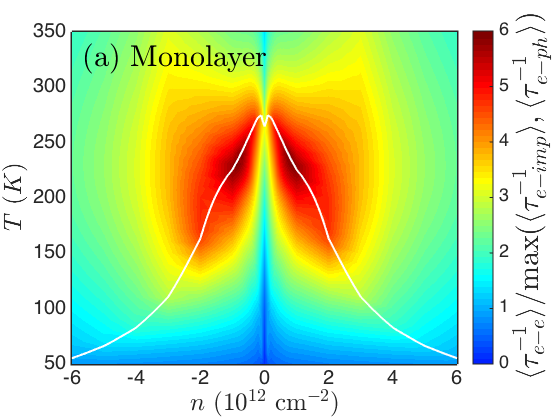} &
\includegraphics[height=!,width=7.5cm] {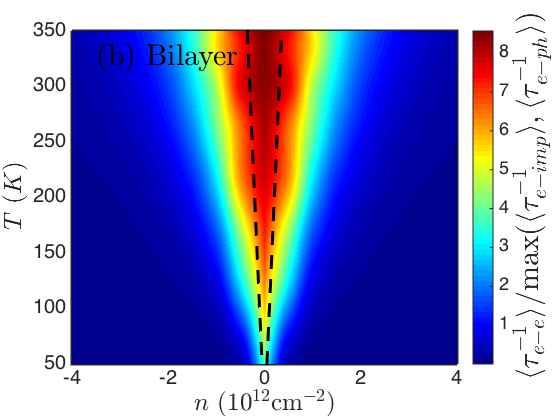} 
\end{array}$
\end{center}
\caption{(Color online) Hydrodynamic window in (a) monolayer graphene and (b) bilayer graphene determined by comparing the carrier-carrier scattering rate to that of impurities and phonons.  For monolayer graphene the ``lung-shape'' is determined by the competition between electron-impurity scattering that dominates at low temperature and carrier density, and electron-phonon scattering that dominates at high temperature and density.  The solid white line traces the points where these two contributions are equal.  For bilayer graphene, we find that the electron-electron scattering is much stronger than in monolayer graphene at the Dirac point, but drops off faster as a function of carrier density.  This yields the ``v-shape'' where electron-electron interactions dominate over electron-phonon interactions at low carrier density, but are weaker at higher carrier density.  The dashed black lines show crossover from the plasma regime that comprises thermal occupation of both electrons and holes to the single carrier regime.  Regardless of amount of disorder, at sufficiently high temperature, bilayer graphene supports both types of hydrodymamics.} \label{Fig2}
\end{figure*}

Our main results can be seen in Fig.~\ref{Fig2}. The color axis shows the ratio of the e-e scattering rate to maximum of the other scattering rates (e-imp and e-ph).  Dark red represents the regions where collisions between carriers dominate the quasiparticle lifetime.  The most striking feature is that the window for hydrodynamics is `lung-shaped' in monolayer graphene, while it is `v-shaped' in bilayer.  The absolute magnitude of the ratio is also larger in bilayer graphene, implying that hydrodynamic effects are stronger.  We now describe some features of the hydrodynamic window.  For monolayer, the e-imp scattering rate diverges at $n\rightarrow 0$ and $T\rightarrow 0$ which explains the reduced hydrodynamic window both close to the Dirac point and at low temperature (this divergence also causes electron-hole puddles which we discuss later).  The white line marks the points in the $(n,T)$ phase space where $\tau_{\rm e-imp} = \tau_{\rm e-ph}$ implying that the crossover to the hydrodynamic window below this line is set entirely by the competition of e-e and e-imp scattering.  Above the white line (i.e.~at higher density and temperature) the emergence of hydrodynamics is determined only by the competition between e-ph and e-e interactions.  The suppression of the window at the Dirac point above the white line is caused by the transverse optical mode at the Brillouin zone boundary, commonly referred to as the $A_1'$ mode~\cite{sohier_phonon-limited_2014}.  At any given temperature, the hydrodynamic window eventually closes with increasing density since both e-e scattering decreases (because of Pauli-blocking and enhanced screening) and acoustic phonon scattering increases.  The lung shape arises because the $A_1'$ phonons peak at the Dirac point and decrease sharply with density, while the acoustic phonons, which are minimal at the Dirac point increase with density. 
Bilayer graphene has a very different hydrodynamic window due to the stronger e-e scattering and weaker e-imp scattering.  Regardless of the amount of disorder, at sufficiently high temperature, e-imp scattering contributes negligibly to the quasiparticle lifetime e.g.~for an estimated $n_{\rm imp}= 9\times 10^{9}~{\rm cm}^{-2}$ for bilayer graphene on h-BN, this is already the case at $T\gtrsim 10~{\rm K}$.  Unlike the monolayer case, there is no high-temperature cutoff to the hydrodynamic window because there is no analog to the $A_1'$ phonon mode~\cite{borysenko_electron-phonon_2011}.  At Dirac point, we find that the acoustic e-ph and e-e scattering rates increase linearly with temperature, but the co-efficient of the e-e scattering rate is much stronger. Hence, we find the hydrodynamic regime in bilayer graphene for the full range of parameters explored in Fig.~\ref{Fig2}. In addition, as indicated by the dashed lines in Fig.~\ref{Fig2}(b) both the plasma hydrodynamic fluid and the single carrier hydrodynamics exist in bilayer graphene. The `v-shape' arises because although e-e scattering peaks at the Dirac point, it decays strongly as a (universal) function of $\varepsilon_F/k_B T$ implying a larger window at higher temperatures.  

\begin{figure*}[!ht]
\begin{center}
$\begin{array}{cc}
\includegraphics[height=!,width=6.5cm] {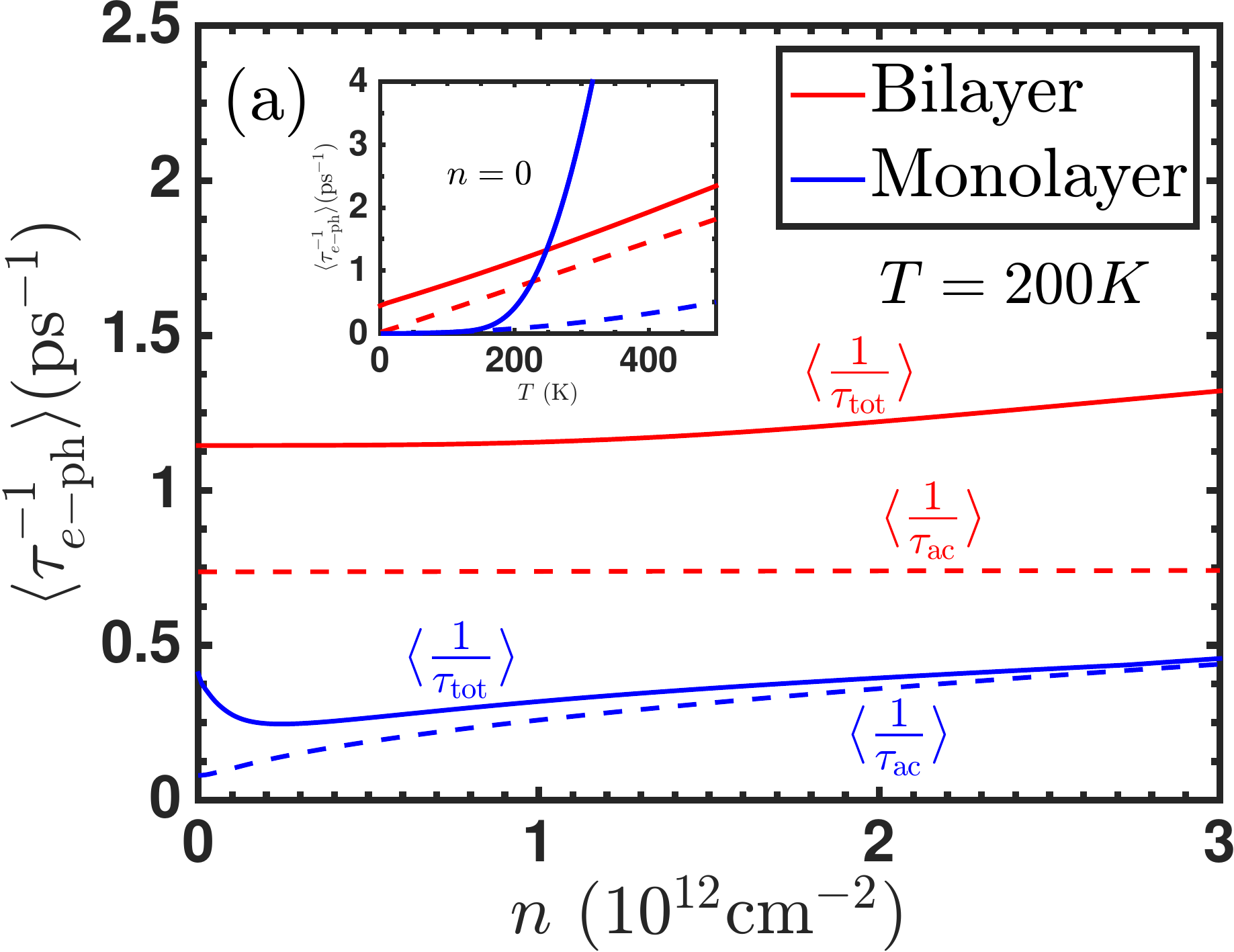}
\includegraphics[height=!,width=6.5cm] {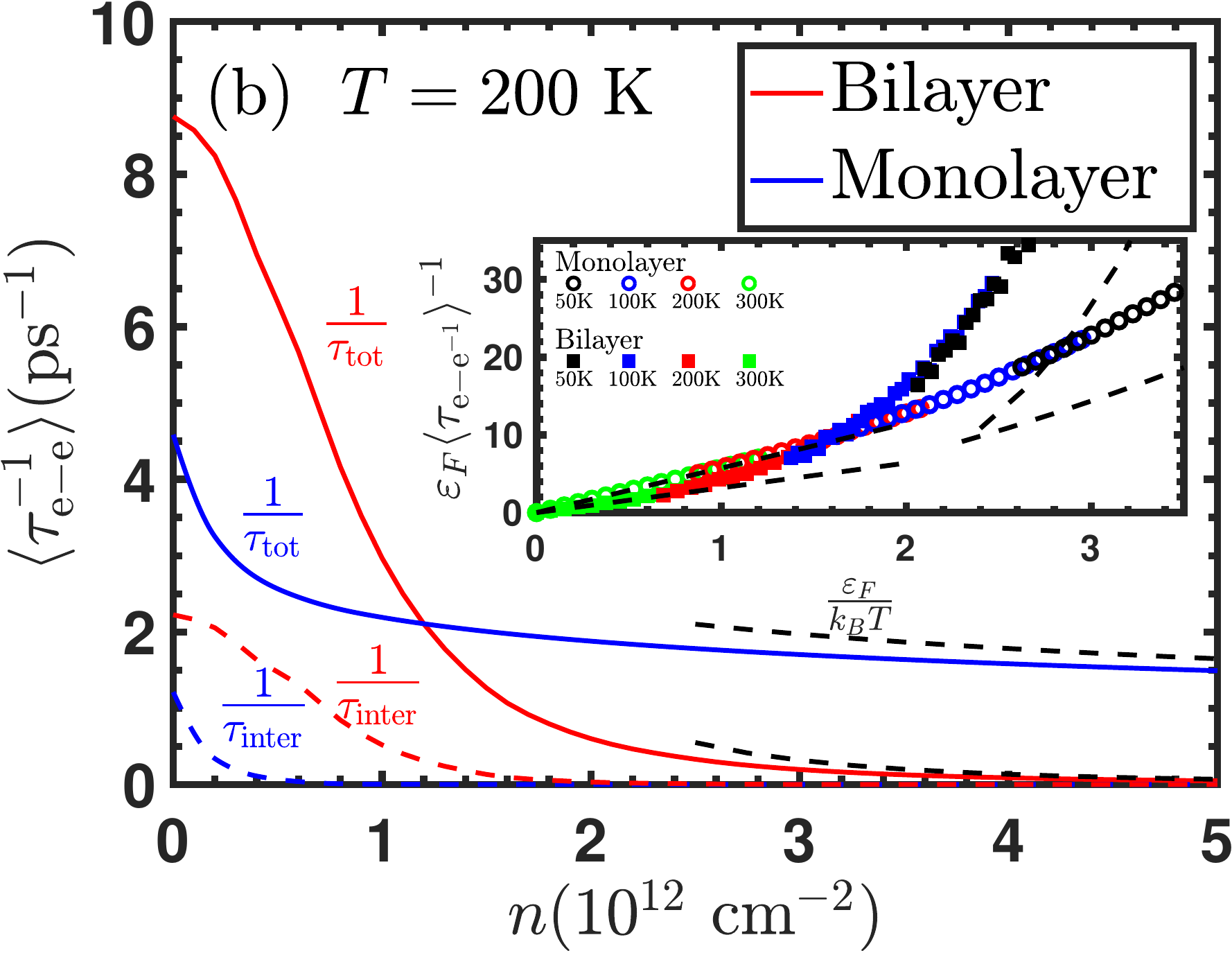} 
\end{array}$
\end{center}
\caption{(color online) (a) Electron-phonon scattering.  Monolayer graphene is mostly acoustic phonons (dashed line) at high density and $A_{1}'$ optical phonon at very low density.  For bilayer, both acoustic (dashed line) and surface polar phonons contribute to the total $\tau_{e-ph}$ (solid lines).  Inset shows results at charge neutrality for the total phonon scattering rate (solid lines) and the acoustic phonon contribution (dashed lines). (b) Electron-electron scattering follows a universal scaling function.  Solid (dashed) lines in the main panel represent the total (interband) e-e scattering rate.  Dashed black lines represent high-density asymptotes of the universal scaling function. Inset shows the collapse of $\epsilon_{\rm F} \tau_{ee} = f(\varepsilon_F/k_B T)$ for different temperatures.  For $\varepsilon_F \ll k_{B} T$, the scaling function is linear for both monolayer and bilayer graphene while for $\varepsilon_F \gg k_{B} T$ (beyond the range shown in the inset) bilayer has a stronger $x^4$ powerlaw compared to the $x^{\sim 2}$ for monolayer.} \label{Fig3} 
\end{figure*}

We now describe our calculations in more detail.  For monolayer graphene, the e-ph scattering is dominated by the in-plane longitudinal and transverse acoustic modes (we use the expressions from Ref.~\cite{sohier_phonon-limited_2014}, but convert the transport scattering rates to quasiparticle rates by removing the ($1 - \cos\theta$) Boltzmann factor before performing the thermal average using Eq.~\ref{th-avrg}).  Our results are displayed in Fig.~\ref{Fig3}(a) (blue curves).  We note that at fixed temperature, electron scattering with $A_1'$ modes (acoustic modes) decreases (increases) with density giving rise to a non-monotonic e-ph scattering rate as a function of density.  
For bilayer the dominant phonons are the acoustic~\cite{borysenko_electron-phonon_2011} and surface polar~\cite{li_electron_2011} modes, which we suitably adapt before doing the thermal average for each mode separately and adding the rates together to obtain the results in Fig.~\ref{Fig3}(a) (red curves).  We note that the total e-ph rate is an almost constant function of density, and increases linear in $T$.  The weak temperature dependence of bilayer phonons compared to monolayer phonons (see inset to Fig.~\ref{Fig3}(a)) illustrates why there is no high temperature cutoff of the bilayer hydrodynamic regime.

Since we find no calculation in the literature of the quasiparticle lifetime in bilayer graphene due to e-e scattering, we calculate it starting from the RPA approximation for the self-energy~\cite{polini_quasiparticle_2014} 
\begin{widetext}
\begin{equation}
\Im m[\Sigma_{\lambda}({\bm{k}},\omega)]=-\int\frac{d^{2}{\bm{q}}}{(2\pi)^{2}}\sum_{\lambda'}\Im m\left[\frac{V(q)}{\epsilon(q,\omega-\varepsilon_{{\bm{k}}-{\bm{q}},\lambda'}+\mu,T)}\right]~{\cal F}_{\lambda\lambda'}(\theta_{{\bm{k}},{\bm{k}}-{\bm{q}}})[n_{{\rm B}}(\omega-\varepsilon_{{\bm{k}}-{\bm{q}},\lambda'}+\mu)+n_{{\rm F}}(-\varepsilon_{{\bm{k}}-{\bm{q}},\lambda'}+\mu)]~, \label{ImSigma}
\end{equation}
\end{widetext}
where $V(q)=2\pi e^2/(\kappa q)$ is the Coulomb potential, $n_{{\rm B}/{\rm F}}(x)\equiv1/[\exp(\beta x)\mp1]$ the Bose (Fermi) distribution function, $\lambda,\lambda'$ are band indices ($\pm$ denote the conduction and valence bands), ${\cal F}_{\lambda\lambda'}(\varphi)= (1+\lambda\lambda'\cos{2\varphi})/2$ the chirality factor, and $\varphi$ the scattering angle. $\varepsilon_{{\bm{k}},\lambda}=\lambda\hbar^2k^2/(2m^\star)$ are parabolic-band single-particle energies and $\epsilon(q, \omega, T) \equiv 1 - V(q) \Pi(q, \omega, T)$ refers to the dynamical RPA dielectric function first calculated in Ref.~\cite{hwang_coulomb_2011}. We show in Fig.~\ref{Fig3}(b) our calculations of $\langle \tau_{\rm ee}^{-1} \rangle$ as a function of density for both monolayer and bilayer graphene.  The primary difference between the two is that for bilayer graphene the scattering rate drops sharply as a function of density whereas for monolayers it drops much more slowly.  Remarkably, we show that after thermal averaging, $\varepsilon_{\rm F} \langle \tau_{\rm ee}^{-1} \rangle^{-1} $ for both monolayer (m) and bilayer (b) graphene is given by a one-parameter function $F_{m,b}(\varepsilon_{\rm F}/k_{\rm B} T)$ (see inset to Fig.~\ref{Fig3}(b)). For monolayers, this is an exact result, while for bilayers, the scaling requires $T \lesssim 2 \alpha^2 (m^* v_{\rm F}^2) \approx 1700\rm{K}$ (here $\alpha$ and $v_F$ are the monolayer fine structure constant and Fermi velocity respectively, and $m^*$ the bilayer effective mass).  This is an especially useful result given that three numerical integrals are required to evaluate e-e scattering at each density and temperature.  We find numerically that the scaling function obeys the following asymptotes: At low density or high temperature, $F_{m}(x \ll 1) \sim 5.7~x$ and $F_{b}(x \ll 1) \sim 3.2~x$, while in the opposite limit $F_{m}(x \gg 1) \sim 2.2~x^{1.7}$ and $F_{b}(x \gg 1) \sim 0.4~x^4$.  Our theoretical finding that the e-e scattering rate at Dirac point for bilayer graphene is linear in temperature i.e.~$\hbar \langle \tau_{ee}^{-1} \rangle = 0.3~k_{B} T$ was recently observed experimentally~\cite{nam_electronhole_2017}.  

\begin{figure}[t!]
\begin{center}
\includegraphics[height=!,width=6.5cm] {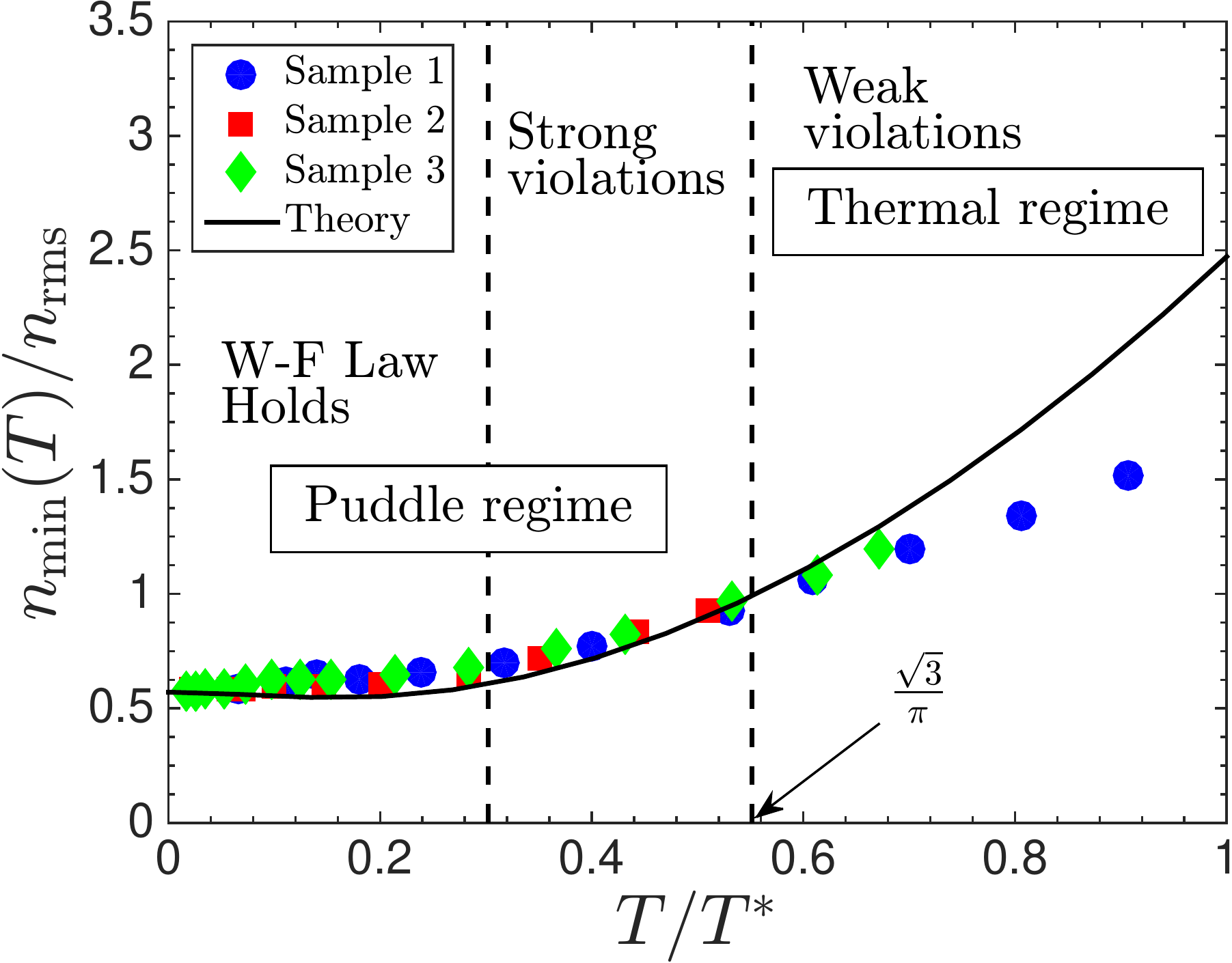} 
\end{center}
\caption{(Color online) Collapse of $n_\mathrm{min}/n_\mathrm{rms}$ as a function of the scaled temperature $T/T^{*}$ where $ k_BT^{*} = \varepsilon_{\rm F}(n = n_{\rm rms})$. Symbols correspond to data extracted from the experiment of Ref.~\cite{crossno_observation_2016}  and the solid line corresponds to the universal disorder limited theory.  The second dashed line marks the temperature when thermally excited carriers equal the disorder induced electron-hole puddles.} \label{Fig4}
\end{figure}
  
So far we have ignored the role of electron-hole puddles. Ref.~\cite{adam_temperature_2010} showed that when impurities were responsible for both scattering electrons and for inducing density fluctuations, the temperature-dependent conductivity followed a one-parameter scaling.  We do this collapse as follows: $n_\mathrm{min}(T)$, a quantity defined in the experimental work, is determined by extrapolating a linear fit of $\log \sigma(n,T)$ down to the minimum conductivity.  The analogous theoretical quantity is $n_\mathrm{rms}$ where Ref.~\cite{adam_self-consistent_2007} showed that $n_\mathrm{min}(T=0) = n_\mathrm{rms}/\sqrt{3} = (k_{B} T^*/\hbar v_{\rm F})^2/\pi$, where $T^*$ is a corresponding disorder temperature scale.  Figure \ref{Fig4} shows the collapse of the experimental $n_{min}/n_\mathrm{\rm rms}$ as a function of experimental $T/T^*$ for all the experimental data of Ref.~\cite{crossno_observation_2016}. This implies that for the experiment, the charge transport is limited by the same mechanism responsible for the puddles.

We calculate this universal disorder limited curve theoretically. We use the transport scattering time calculated using the RPA-effective medium conductivity~\cite{hwang_screening-induced_2009,rossi_effective_2009} as a function of $n$ and $T$ and extract the effective density at Dirac point $n_{min}(T)$ using the same procedure described above. Our theory shows excellent agreement with experiment over a wide range, with weak deviations occurring only at high temperatures $T/T^{*} \gtrsim 0.55$.  (This is when thermal excited carriers outnumber the puddle carriers induced by impurities i.e.~$n_{e}(0,T_c) + n_{h}(0,T_c) = n_\mathrm{rms}$, where $T_c =\sqrt{3} T^{*}/\pi \approx 0.55~ T^*$). 
This confirms that electrical conductivity at temperatures below this is completely understood within a model considering only charged impurity scattering. 

Our results allow us to speculate regarding the recent experiments.  The Harvard experiments observed a strong violation of the Wiedemann-Franz law (with Lorenz ratios as high as $22$).  This was interpreted as a signature of the theoretically predicted Dirac fluid~\cite{fritz_quantum_2008,muller_quantum-critical_2008} where the thermal conductivity diverges and the electrical conductivity approaches a universal value of $\sigma_0 = 4/\pi$.  This would imply a non-monotonic temperature dependence as the plasma evolved from the disorder dominated puddle regime to the ideal Dirac fluid regime.  Not only does the experimental electrical conductivity increase monotonically, but the temperature dependence quantitatively agrees with the disorder limited theory (in addition, the absolute value of the conductivity is a factor of 5 larger than $\sigma_0$).  Therefore the violations of Wiedemann-Franz seen in experiment arise from the thermal rather than the charge transport sector (where this enhancement occurs because in the plasma regime carrier-carrier collisions are unable to relax thermal currents).  This also implies that if one moves out of the plasma regime (i.e.~within the ``lung-region'' in Fig.~\ref{Fig2}), then there should be weak violation of the Wiedemann-Franz law with a ${\it reduced}$ Lorenz ratio, because in the single-carrier regime charge currents are unaffected by carrier-carrier collisions, while thermal currents can be degraded~\cite{principi_violation_2015}.  Our results show that the hydrodynamic regime is stronger in bilayer graphene, becoming increasingly stronger at higher temperature.  However, we emphasize that deciding when the system is hydrodynamic and when there is a violation of the Wiedemann-Franz law are two separate questions, and one can have one without the other~\cite{xie_transport_2016}.  In particular, the argument for the conservation of thermal conductivity at the Dirac point in monolayer graphene does not hold in bilayer graphene.  So although bilayer graphene is {\it more} hydrodynamic than monolayer graphene, we do not expect a strong violation in the Widemann-Franz law.       

Turning to the Manchester experiments, there is good qualitative agreement with the shapes of hydrodynamic window shown in Fig.~\ref{Fig2} (i.e.~a `lung' for monolayer and a `v-shape' for bilayer) and those seen experimentally.  We note that no whirlpools were observed in the plasma hydrodynamic regime (dashed lines in Fig.~\ref{Fig2}b), possibly because of the presence of electrons and holes moving in opposite directions.  In particular, our finding that the monolayer hydrodynamic window is cutoff by phonons at high temperature, but not in bilayer graphene is consistent with the experimental observations.  The details of the `lung-shape' in monolayer graphene is somewhat different, but our conclusion that the low temperature boundary is set by impurities and the high temperature boundary is set by phonons should remain robust regardless of the impurity and phonon models used (e.g.~it seems that the ratio of $A_1'$ to acoustic phonon modes is stronger in the experiment than in the best available models~\cite{sohier_phonon-limited_2014}).  Although the bilayer e-e scattering rate is larger at higher temperature, giving a wider window in carrier density where hydrodynamics can be observed, the viscosity is lower at higher temperature making the whirlpools more difficult to observe.  Again, just like for the violation of the Widemann-Franz law, there is not a perfect correlation between the hydrodynamic regime and observations of negative vicinity resistance.

{\it Acknowledgement:} It is a pleasure to thank Sankar Das Sarma, Matthew Foster, Ben Yu-Kuang Hu, Philip Kim, Andrew Lucas, 
Boris Narozhny, Marco Polini, Subir Sachdev, and Giovanni Vignale for valuable discussions, and acknowledge the support of the National Research Foundation of Singapore under its Fellowship program (NRF-NRFF2012-01).

{\it Note Added:} After completion of this work, we noticed on arXiv a useful review article on graphene hydrodynamics~\cite{lucas_hydrodynamics_2017} that provides a broader context for the relevance of our work.


\begin{thebibliography}{30}%
\makeatletter
\providecommand \@ifxundefined [1]{%
 \@ifx{#1\undefined}
}%
\providecommand \@ifnum [1]{%
 \ifnum #1\expandafter \@firstoftwo
 \else \expandafter \@secondoftwo
 \fi
}%
\providecommand \@ifx [1]{%
 \ifx #1\expandafter \@firstoftwo
 \else \expandafter \@secondoftwo
 \fi
}%
\providecommand \natexlab [1]{#1}%
\providecommand \enquote  [1]{``#1''}%
\providecommand \bibnamefont  [1]{#1}%
\providecommand \bibfnamefont [1]{#1}%
\providecommand \citenamefont [1]{#1}%
\providecommand \href@noop [0]{\@secondoftwo}%
\providecommand \href [0]{\begingroup \@sanitize@url \@href}%
\providecommand \@href[1]{\@@startlink{#1}\@@href}%
\providecommand \@@href[1]{\endgroup#1\@@endlink}%
\providecommand \@sanitize@url [0]{\catcode `\\12\catcode `\$12\catcode
  `\&12\catcode `\#12\catcode `\^12\catcode `\_12\catcode `\%12\relax}%
\providecommand \@@startlink[1]{}%
\providecommand \@@endlink[0]{}%
\providecommand \url  [0]{\begingroup\@sanitize@url \@url }%
\providecommand \@url [1]{\endgroup\@href {#1}{\urlprefix }}%
\providecommand \urlprefix  [0]{URL }%
\providecommand \Eprint [0]{\href }%
\providecommand \doibase [0]{http://dx.doi.org/}%
\providecommand \selectlanguage [0]{\@gobble}%
\providecommand \bibinfo  [0]{\@secondoftwo}%
\providecommand \bibfield  [0]{\@secondoftwo}%
\providecommand \translation [1]{[#1]}%
\providecommand \BibitemOpen [0]{}%
\providecommand \bibitemStop [0]{}%
\providecommand \bibitemNoStop [0]{.\EOS\space}%
\providecommand \EOS [0]{\spacefactor3000\relax}%
\providecommand \BibitemShut  [1]{\csname bibitem#1\endcsname}%
\let\auto@bib@innerbib\@empty
\bibitem [{\citenamefont {Landau}\ and\ \citenamefont
  {Lifshitz}(1981)}]{landau_course_1981}%
  \BibitemOpen
  \bibfield  {author} {\bibinfo {author} {\bibfnamefont {L.~D.}\ \bibnamefont
  {Landau}}\ and\ \bibinfo {author} {\bibfnamefont {E.~M.}\ \bibnamefont
  {Lifshitz}},\ }\href@noop {} {\emph {\bibinfo {title} {Course of
  {Theoretical} {Physics}: {Physical} {Kinetics} : by {E}.{M}. {Lifshitz} and
  {L}.{P}. {Pitaevskii}}}}\ (\bibinfo {year} {1981})\BibitemShut {NoStop}%
\bibitem [{\citenamefont {Gurzhi}(1968)}]{gurzhi_hydrodynamic_1968}%
  \BibitemOpen
  \bibfield  {author} {\bibinfo {author} {\bibfnamefont {R.~N.}\ \bibnamefont
  {Gurzhi}},\ }\href {\doibase 10.1070/PU1968v011n02ABEH003815} {\bibfield
  {journal} {\bibinfo  {journal} {Soviet Physics Uspekhi}\ }\textbf {\bibinfo
  {volume} {11}},\ \bibinfo {pages} {255} (\bibinfo {year} {1968})}\BibitemShut
  {NoStop}%
\bibitem [{\citenamefont {Govorov}\ and\ \citenamefont
  {Heremans}(2004)}]{govorov_hydrodynamic_2004}%
  \BibitemOpen
  \bibfield  {author} {\bibinfo {author} {\bibfnamefont {A.~O.}\ \bibnamefont
  {Govorov}}\ and\ \bibinfo {author} {\bibfnamefont {J.~J.}\ \bibnamefont
  {Heremans}},\ }\href {\doibase 10.1103/PhysRevLett.92.026803} {\bibfield
  {journal} {\bibinfo  {journal} {Physical Review Letters}\ }\textbf {\bibinfo
  {volume} {92}},\ \bibinfo {pages} {026803} (\bibinfo {year}
  {2004})}\BibitemShut {NoStop}%
\bibitem [{\citenamefont {Dyakonov}(1993)}]{dyakonov_shallow_1993}%
  \BibitemOpen
  \bibfield  {author} {\bibinfo {author} {\bibfnamefont {M.}~\bibnamefont
  {Dyakonov}},\ }\href {\doibase 10.1103/PhysRevLett.71.2465} {\bibfield
  {journal} {\bibinfo  {journal} {Physical Review Letters}\ }\textbf {\bibinfo
  {volume} {71}},\ \bibinfo {pages} {2465} (\bibinfo {year}
  {1993})}\BibitemShut {NoStop}%
\bibitem [{\citenamefont {Avron}\ \emph {et~al.}(1995)\citenamefont {Avron},
  \citenamefont {Seiler},\ and\ \citenamefont {Zograf}}]{avron_viscosity_1995}%
  \BibitemOpen
  \bibfield  {author} {\bibinfo {author} {\bibfnamefont {J.~E.}\ \bibnamefont
  {Avron}}, \bibinfo {author} {\bibfnamefont {R.}~\bibnamefont {Seiler}}, \
  and\ \bibinfo {author} {\bibfnamefont {P.~G.}\ \bibnamefont {Zograf}},\
  }\href {\doibase 10.1103/PhysRevLett.75.697} {\bibfield  {journal} {\bibinfo
  {journal} {Physical Review Letters}\ }\textbf {\bibinfo {volume} {75}},\
  \bibinfo {pages} {697} (\bibinfo {year} {1995})}\BibitemShut {NoStop}%
\bibitem [{\citenamefont {Alekseev}(2016)}]{alekseev_negative_2016}%
  \BibitemOpen
  \bibfield  {author} {\bibinfo {author} {\bibfnamefont {P.}~\bibnamefont
  {Alekseev}},\ }\href {\doibase 10.1103/PhysRevLett.117.166601} {\bibfield
  {journal} {\bibinfo  {journal} {Physical Review Letters}\ }\textbf {\bibinfo
  {volume} {117}},\ \bibinfo {pages} {166601} (\bibinfo {year}
  {2016})}\BibitemShut {NoStop}%
\bibitem [{\citenamefont {Guo}\ \emph {et~al.}(2017)\citenamefont {Guo},
  \citenamefont {Ilseven}, \citenamefont {Falkovich},\ and\ \citenamefont
  {Levitov}}]{guo_higher-than-ballistic_2017}%
  \BibitemOpen
  \bibfield  {author} {\bibinfo {author} {\bibfnamefont {H.}~\bibnamefont
  {Guo}}, \bibinfo {author} {\bibfnamefont {E.}~\bibnamefont {Ilseven}},
  \bibinfo {author} {\bibfnamefont {G.}~\bibnamefont {Falkovich}}, \ and\
  \bibinfo {author} {\bibfnamefont {L.~S.}\ \bibnamefont {Levitov}},\ }\href
  {\doibase 10.1073/pnas.1612181114} {\bibfield  {journal}{\bibinfo  {journal}
  {Proceedings of the National Academy of Sciences}\ } {\ \textbf {\bibinfo {volume} {114}},\
  \bibinfo {pages} {3068} (\bibinfo {year} {2017})}}\BibitemShut {NoStop}%
\bibitem [{\citenamefont {de~Jong}\ and\ \citenamefont
  {Molenkamp}(1995)}]{de_jong_hydrodynamic_1995}%
  \BibitemOpen
  \bibfield  {author} {\bibinfo {author} {\bibfnamefont {M.~J.~M.}\
  \bibnamefont {de~Jong}}\ and\ \bibinfo {author} {\bibfnamefont {L.~W.}\
  \bibnamefont {Molenkamp}},\ }\href {\doibase 10.1103/PhysRevB.51.13389}
  {\bibfield  {journal} {\bibinfo  {journal} {Physical Review B}\ }\textbf
  {\bibinfo {volume} {51}},\ \bibinfo {pages} {13389} (\bibinfo {year}
  {1995})}\BibitemShut {NoStop}%
\bibitem [{\citenamefont {Bandurin}\ \emph {et~al.}(2016)\citenamefont
  {Bandurin}, \citenamefont {Torre}, \citenamefont {Kumar}, \citenamefont
  {Shalom}, \citenamefont {Tomadin}, \citenamefont {Principi}, \citenamefont
  {Auton}, \citenamefont {Khestanova}, \citenamefont {Novoselov}, \citenamefont
  {Grigorieva}, \citenamefont {Ponomarenko}, \citenamefont {Geim},\ and\
  \citenamefont {Polini}}]{bandurin_negative_2016}%
  \BibitemOpen
  \bibfield  {author} {\bibinfo {author} {\bibfnamefont {D.~A.}\ \bibnamefont
  {Bandurin}}, \bibinfo {author} {\bibfnamefont {I.}~\bibnamefont {Torre}},
  \bibinfo {author} {\bibfnamefont {R.~K.}\ \bibnamefont {Kumar}}, \bibinfo
  {author} {\bibfnamefont {M.~B.}\ \bibnamefont {Shalom}}, \bibinfo {author}
  {\bibfnamefont {A.}~\bibnamefont {Tomadin}}, \bibinfo {author} {\bibfnamefont
  {A.}~\bibnamefont {Principi}}, \bibinfo {author} {\bibfnamefont {G.~H.}\
  \bibnamefont {Auton}}, \bibinfo {author} {\bibfnamefont {E.}~\bibnamefont
  {Khestanova}}, \bibinfo {author} {\bibfnamefont {K.~S.}\ \bibnamefont
  {Novoselov}}, \bibinfo {author} {\bibfnamefont {I.~V.}\ \bibnamefont
  {Grigorieva}}, \bibinfo {author} {\bibfnamefont {L.~A.}\ \bibnamefont
  {Ponomarenko}}, \bibinfo {author} {\bibfnamefont {A.~K.}\ \bibnamefont
  {Geim}}, \ and\ \bibinfo {author} {\bibfnamefont {M.}~\bibnamefont
  {Polini}},\ }\href {\doibase 10.1126/science.aad0201} {\bibfield  {journal}
  {\bibinfo  {journal} {Science}\ }\textbf {\bibinfo {volume} {351}},\ \bibinfo
  {pages} {1055} (\bibinfo {year} {2016})}\BibitemShut {NoStop}%
\bibitem [{\citenamefont {Crossno}\ \emph {et~al.}(2016)\citenamefont
  {Crossno}, \citenamefont {Shi}, \citenamefont {Wang}, \citenamefont {Liu},
  \citenamefont {Harzheim}, \citenamefont {Lucas}, \citenamefont {Sachdev},
  \citenamefont {Kim}, \citenamefont {Taniguchi}, \citenamefont {Watanabe},
  \citenamefont {Ohki},\ and\ \citenamefont {Fong}}]{crossno_observation_2016}%
  \BibitemOpen
  \bibfield  {author} {\bibinfo {author} {\bibfnamefont {J.}~\bibnamefont
  {Crossno}}, \bibinfo {author} {\bibfnamefont {J.~K.}\ \bibnamefont {Shi}},
  \bibinfo {author} {\bibfnamefont {K.}~\bibnamefont {Wang}}, \bibinfo {author}
  {\bibfnamefont {X.}~\bibnamefont {Liu}}, \bibinfo {author} {\bibfnamefont
  {A.}~\bibnamefont {Harzheim}}, \bibinfo {author} {\bibfnamefont
  {A.}~\bibnamefont {Lucas}}, \bibinfo {author} {\bibfnamefont
  {S.}~\bibnamefont {Sachdev}}, \bibinfo {author} {\bibfnamefont
  {P.}~\bibnamefont {Kim}}, \bibinfo {author} {\bibfnamefont {T.}~\bibnamefont
  {Taniguchi}}, \bibinfo {author} {\bibfnamefont {K.}~\bibnamefont {Watanabe}},
  \bibinfo {author} {\bibfnamefont {T.~A.}\ \bibnamefont {Ohki}}, \ and\
  \bibinfo {author} {\bibfnamefont {K.~C.}\ \bibnamefont {Fong}},\ }\href
  {\doibase 10.1126/science.aad0343} {\bibfield  {journal} {\bibinfo  {journal}
  {Science}\ }\textbf {\bibinfo {volume} {351}},\ \bibinfo {pages} {1058}
  (\bibinfo {year} {2016})}\BibitemShut {NoStop}%
\bibitem [{\citenamefont {Nam}\ \emph {et~al.}(2017)\citenamefont {Nam},
  \citenamefont {Ki}, \citenamefont {Soler-Delgado},\ and\ \citenamefont
  {Morpurgo}}]{nam_electronhole_2017}%
  \BibitemOpen
  \bibfield  {author} {\bibinfo {author} {\bibfnamefont {Y.}~\bibnamefont
  {Nam}}, \bibinfo {author} {\bibfnamefont {D.-K.}\ \bibnamefont {Ki}},
  \bibinfo {author} {\bibfnamefont {D.}~\bibnamefont {Soler-Delgado}}, \ and\
  \bibinfo {author} {\bibfnamefont {A.~F.}\ \bibnamefont {Morpurgo}},\ }\href
  {\doibase 10.1038/nphys4218} {\bibfield  {journal} {\bibinfo  {journal}
  {Nature Physics (in press)}\ ,\ \bibinfo {pages} {available online at DOI 10.1038/nphys4218}} (\bibinfo {year}
  {2017})}\BibitemShut {NoStop}%
\bibitem [{\citenamefont {Moll}\ \emph {et~al.}(2016)\citenamefont {Moll},
  \citenamefont {Kushwaha}, \citenamefont {Nandi}, \citenamefont {Schmidt},\
  and\ \citenamefont {Mackenzie}}]{moll_evidence_2016}%
  \BibitemOpen
  \bibfield  {author} {\bibinfo {author} {\bibfnamefont {P.~J.~W.}\
  \bibnamefont {Moll}}, \bibinfo {author} {\bibfnamefont {P.}~\bibnamefont
  {Kushwaha}}, \bibinfo {author} {\bibfnamefont {N.}~\bibnamefont {Nandi}},
  \bibinfo {author} {\bibfnamefont {B.}~\bibnamefont {Schmidt}}, \ and\
  \bibinfo {author} {\bibfnamefont {A.~P.}\ \bibnamefont {Mackenzie}},\ }\href
  {\doibase 10.1126/science.aac8385} {\bibfield  {journal} {\bibinfo  {journal}
  {Science}\ }\textbf {\bibinfo {volume} {351}},\ \bibinfo {pages} {1061}
  (\bibinfo {year} {2016})}\BibitemShut {NoStop}%
\bibitem [{\citenamefont {Torre}\ \emph {et~al.}(2015)\citenamefont {Torre},
  \citenamefont {Tomadin}, \citenamefont {Geim},\ and\ \citenamefont
  {Polini}}]{torre_nonlocal_2015}%
  \BibitemOpen
  \bibfield  {author} {\bibinfo {author} {\bibfnamefont {I.}~\bibnamefont
  {Torre}}, \bibinfo {author} {\bibfnamefont {A.}~\bibnamefont {Tomadin}},
  \bibinfo {author} {\bibfnamefont {A.~K.}\ \bibnamefont {Geim}}, \ and\
  \bibinfo {author} {\bibfnamefont {M.}~\bibnamefont {Polini}},\ }\href
  {\doibase 10.1103/PhysRevB.92.165433} {\bibfield  {journal} {\bibinfo
  {journal} {Physical Review B}\ }\textbf {\bibinfo {volume} {92}},\ \bibinfo
  {pages} {165433} (\bibinfo {year} {2015})}\BibitemShut {NoStop}%
\bibitem [{\citenamefont {Narozhny}\ \emph {et~al.}(2015)\citenamefont
  {Narozhny}, \citenamefont {Gornyi}, \citenamefont {Titov}, \citenamefont
  {Schütt},\ and\ \citenamefont {Mirlin}}]{narozhny_hydrodynamics_2015}%
  \BibitemOpen
  \bibfield  {author} {\bibinfo {author} {\bibfnamefont {B.~N.}\ \bibnamefont
  {Narozhny}}, \bibinfo {author} {\bibfnamefont {I.~V.}\ \bibnamefont
  {Gornyi}}, \bibinfo {author} {\bibfnamefont {M.}~\bibnamefont {Titov}},
  \bibinfo {author} {\bibfnamefont {M.}~\bibnamefont {Sch{\"u}tt}}, \ and\
  \bibinfo {author} {\bibfnamefont {A.~D.}\ \bibnamefont {Mirlin}},\ }\href
  {\doibase 10.1103/PhysRevB.91.035414} {\bibfield  {journal} {\bibinfo
  {journal} {Physical Review B}\ }\textbf {\bibinfo {volume} {91}},\ \bibinfo
  {pages} {035414} (\bibinfo {year} {2015})}\BibitemShut {NoStop}%
\bibitem [{\citenamefont {Lucas}\ \emph {et~al.}(2016)\citenamefont {Lucas},
  \citenamefont {Crossno}, \citenamefont {Fong}, \citenamefont {Kim},\ and\
  \citenamefont {Sachdev}}]{lucas_transport_2016}%
  \BibitemOpen
  \bibfield  {author} {\bibinfo {author} {\bibfnamefont {A.}~\bibnamefont
  {Lucas}}, \bibinfo {author} {\bibfnamefont {J.}~\bibnamefont {Crossno}},
  \bibinfo {author} {\bibfnamefont {K.~C.}\ \bibnamefont {Fong}}, \bibinfo
  {author} {\bibfnamefont {P.}~\bibnamefont {Kim}}, \ and\ \bibinfo {author}
  {\bibfnamefont {S.}~\bibnamefont {Sachdev}},\ }\href {\doibase
  10.1103/PhysRevB.93.075426} {\bibfield  {journal} {\bibinfo  {journal}
  {Physical Review B}\ }\textbf {\bibinfo {volume} {93}},\ \bibinfo {pages}
  {075426} (\bibinfo {year} {2016})}\BibitemShut {NoStop}%
\bibitem [{\citenamefont {Sch\"{u}tt}\ \emph {et~al.}(2011)\citenamefont
  {Sch\"{u}tt}, \citenamefont {Ostrovsky}, \citenamefont {Gornyi},\ and\
  \citenamefont {Mirlin}}]{schutt_coulomb_2011}%
  \BibitemOpen
  \bibfield  {author} {\bibinfo {author} {\bibfnamefont {M.}~\bibnamefont
  {Sch\"{u}tt}}, \bibinfo {author} {\bibfnamefont {P.~M.}\ \bibnamefont
  {Ostrovsky}}, \bibinfo {author} {\bibfnamefont {I.~V.}\ \bibnamefont
  {Gornyi}}, \ and\ \bibinfo {author} {\bibfnamefont {A.~D.}\ \bibnamefont
  {Mirlin}},\ }\href {\doibase 10.1103/PhysRevB.83.155441} {\bibfield
  {journal} {\bibinfo  {journal} {Physical Review B}\ }\textbf {\bibinfo
  {volume} {83}},\ \bibinfo {pages} {155441} (\bibinfo {year}
  {2011})}\BibitemShut {NoStop}%
\bibitem [{\citenamefont {Adam}\ and\ \citenamefont
  {Stiles}(2010)}]{adam_temperature_2010}%
  \BibitemOpen
  \bibfield  {author} {\bibinfo {author} {\bibfnamefont {S.}~\bibnamefont
  {Adam}}\ and\ \bibinfo {author} {\bibfnamefont {M.~D.}\ \bibnamefont
  {Stiles}},\ }\href {http://arxiv.org/abs/0912.1606} {\bibfield  {journal}
  {\bibinfo  {journal} {Physical Review B}\ }\textbf {\bibinfo {volume} {82}}
  (\bibinfo {year} {2010})},\ \bibinfo {note} {arXiv: 0912.1606}\BibitemShut
  {NoStop}%
\bibitem [{\citenamefont {Polini}\ and\ \citenamefont
  {Vignale}(2014)}]{polini_quasiparticle_2014}%
  \BibitemOpen
  \bibfield  {author} {\bibinfo {author} {\bibfnamefont {M.}~\bibnamefont
  {Polini}}\ and\ \bibinfo {author} {\bibfnamefont {G.}~\bibnamefont
  {Vignale}},\ }\href {http://arxiv.org/abs/1404.5728} {\bibfield  {journal}
  {\bibinfo  {journal} {arXiv:1404.5728 [cond-mat]}\ } (\bibinfo {year}
  {2014})}\BibitemShut {NoStop}%
\bibitem [{\citenamefont {Hwang}\ and\ \citenamefont
  {Das~Sarma}(2008)}]{hwang_single-particle_2008}%
  \BibitemOpen
  \bibfield  {author} {\bibinfo {author} {\bibfnamefont {E.~H.}\ \bibnamefont
  {Hwang}}\ and\ \bibinfo {author} {\bibfnamefont {S.}~\bibnamefont
  {Das~Sarma}},\ }\href {\doibase 10.1103/PhysRevB.77.195412} {\bibfield
  {journal} {\bibinfo  {journal} {Physical Review B}\ }\textbf {\bibinfo
  {volume} {77}},\ \bibinfo {pages} {195412} (\bibinfo {year}
  {2008})}\BibitemShut {NoStop}%
\bibitem [{\citenamefont {Sohier}\ \emph {et~al.}(2014)\citenamefont {Sohier},
  \citenamefont {Calandra}, \citenamefont {Park}, \citenamefont {Bonini},
  \citenamefont {Marzari},\ and\ \citenamefont
  {Mauri}}]{sohier_phonon-limited_2014}%
  \BibitemOpen
  \bibfield  {author} {\bibinfo {author} {\bibfnamefont {T.}~\bibnamefont
  {Sohier}}, \bibinfo {author} {\bibfnamefont {M.}~\bibnamefont {Calandra}},
  \bibinfo {author} {\bibfnamefont {C.-H.}\ \bibnamefont {Park}}, \bibinfo
  {author} {\bibfnamefont {N.}~\bibnamefont {Bonini}}, \bibinfo {author}
  {\bibfnamefont {N.}~\bibnamefont {Marzari}}, \ and\ \bibinfo {author}
  {\bibfnamefont {F.}~\bibnamefont {Mauri}},\ }\href {\doibase
  10.1103/PhysRevB.90.125414} {\bibfield  {journal} {\bibinfo  {journal}
  {Physical Review B}\ }\textbf {\bibinfo {volume} {90}},\ \bibinfo {pages}
  {125414} (\bibinfo {year} {2014})}\BibitemShut {NoStop}%
\bibitem [{\citenamefont {Borysenko}\ \emph {et~al.}(2011)\citenamefont
  {Borysenko}, \citenamefont {Mullen}, \citenamefont {Li}, \citenamefont
  {Semenov}, \citenamefont {Zavada}, \citenamefont {Nardelli},\ and\
  \citenamefont {Kim}}]{borysenko_electron-phonon_2011}%
  \BibitemOpen
  \bibfield  {author} {\bibinfo {author} {\bibfnamefont {K.~M.}\ \bibnamefont
  {Borysenko}}, \bibinfo {author} {\bibfnamefont {J.~T.}\ \bibnamefont
  {Mullen}}, \bibinfo {author} {\bibfnamefont {X.}~\bibnamefont {Li}}, \bibinfo
  {author} {\bibfnamefont {Y.~G.}\ \bibnamefont {Semenov}}, \bibinfo {author}
  {\bibfnamefont {J.~M.}\ \bibnamefont {Zavada}}, \bibinfo {author}
  {\bibfnamefont {M.~B.}\ \bibnamefont {Nardelli}}, \ and\ \bibinfo {author}
  {\bibfnamefont {K.~W.}\ \bibnamefont {Kim}},\ }\href {\doibase
  10.1103/PhysRevB.83.161402} {\bibfield  {journal} {\bibinfo  {journal}
  {Physical Review B}\ }\textbf {\bibinfo {volume} {83}},\ \bibinfo {pages}
  {161402} (\bibinfo {year} {2011})}\BibitemShut {NoStop}%
\bibitem [{\citenamefont {Li}\ \emph {et~al.}(2011)\citenamefont {Li},
  \citenamefont {Borysenko}, \citenamefont {Nardelli},\ and\ \citenamefont
  {Kim}}]{li_electron_2011}%
  \BibitemOpen
  \bibfield  {author} {\bibinfo {author} {\bibfnamefont {X.}~\bibnamefont
  {Li}}, \bibinfo {author} {\bibfnamefont {K.~M.}\ \bibnamefont {Borysenko}},
  \bibinfo {author} {\bibfnamefont {M.~B.}\ \bibnamefont {Nardelli}}, \ and\
  \bibinfo {author} {\bibfnamefont {K.~W.}\ \bibnamefont {Kim}},\ }\href
  {\doibase 10.1103/PhysRevB.84.195453} {\bibfield  {journal} {\bibinfo
  {journal} {Physical Review B}\ }\textbf {\bibinfo {volume} {84}},\ \bibinfo
  {pages} {195453} (\bibinfo {year} {2011})}\BibitemShut {NoStop}%
\bibitem [{\citenamefont {Hwang}\ \emph {et~al.}(2011)\citenamefont {Hwang},
  \citenamefont {Sensarma},\ and\ \citenamefont
  {Das~Sarma}}]{hwang_coulomb_2011}%
  \BibitemOpen
  \bibfield  {author} {\bibinfo {author} {\bibfnamefont {E.~H.}\ \bibnamefont
  {Hwang}}, \bibinfo {author} {\bibfnamefont {R.}~\bibnamefont {Sensarma}}, \
  and\ \bibinfo {author} {\bibfnamefont {S.}~\bibnamefont {Das~Sarma}},\ }\href
  {\doibase 10.1103/PhysRevB.84.245441} {\bibfield  {journal} {\bibinfo
  {journal} {Physical Review B}\ }\textbf {\bibinfo {volume} {84}},\ \bibinfo
  {pages} {245441} (\bibinfo {year} {2011})}\BibitemShut {NoStop}%
\bibitem [{\citenamefont {Adam}\ \emph {et~al.}(2007)\citenamefont {Adam},
  \citenamefont {Hwang}, \citenamefont {Galitski},\ and\ \citenamefont
  {Sarma}}]{adam_self-consistent_2007}%
  \BibitemOpen
  \bibfield  {author} {\bibinfo {author} {\bibfnamefont {S.}~\bibnamefont
  {Adam}}, \bibinfo {author} {\bibfnamefont {E.~H.}\ \bibnamefont {Hwang}},
  \bibinfo {author} {\bibfnamefont {V.~M.}\ \bibnamefont {Galitski}}, \ and\
  \bibinfo {author} {\bibfnamefont {S.~Das.}\ \bibnamefont {Sarma}},\ }\href
  {\doibase 10.1073/pnas.0704772104} {\bibfield  {journal} {\bibinfo  {journal}
  {Proceedings of the National Academy of Sciences}\ }\textbf {\bibinfo
  {volume} {104}},\ \bibinfo {pages} {18392} (\bibinfo {year}
  {2007})}\BibitemShut {NoStop}%
\bibitem [{\citenamefont {Hwang}\ and\ \citenamefont
  {Das~Sarma}(2009)}]{hwang_screening-induced_2009}%
  \BibitemOpen
  \bibfield  {author} {\bibinfo {author} {\bibfnamefont {E.~H.}\ \bibnamefont
  {Hwang}}\ and\ \bibinfo {author} {\bibfnamefont {S.}~\bibnamefont
  {Das~Sarma}},\ }\href {\doibase 10.1103/PhysRevB.79.165404} {\bibfield
  {journal} {\bibinfo  {journal} {Physical Review B}\ }\textbf {\bibinfo
  {volume} {79}},\ \bibinfo {pages} {165404} (\bibinfo {year}
  {2009})}\BibitemShut {NoStop}%
\bibitem [{\citenamefont {Rossi}\ \emph {et~al.}(2009)\citenamefont {Rossi},
  \citenamefont {Adam},\ and\ \citenamefont
  {Das~Sarma}}]{rossi_effective_2009}%
  \BibitemOpen
  \bibfield  {author} {\bibinfo {author} {\bibfnamefont {E.}~\bibnamefont
  {Rossi}}, \bibinfo {author} {\bibfnamefont {S.}~\bibnamefont {Adam}}, \ and\
  \bibinfo {author} {\bibfnamefont {S.}~\bibnamefont {Das~Sarma}},\ }\href
  {\doibase 10.1103/PhysRevB.79.245423} {\bibfield  {journal} {\bibinfo
  {journal} {Physical Review B}\ }\textbf {\bibinfo {volume} {79}},\ \bibinfo
  {pages} {245423} (\bibinfo {year} {2009})}\BibitemShut {NoStop}%
\bibitem [{\citenamefont {Fritz}\ \emph {et~al.}(2008)\citenamefont {Fritz},
  \citenamefont {Schmalian}, \citenamefont {Müller},\ and\ \citenamefont
  {Sachdev}}]{fritz_quantum_2008}%
  \BibitemOpen
  \bibfield  {author} {\bibinfo {author} {\bibfnamefont {L.}~\bibnamefont
  {Fritz}}, \bibinfo {author} {\bibfnamefont {J.}~\bibnamefont {Schmalian}},
  \bibinfo {author} {\bibfnamefont {M.}~\bibnamefont {Müller}}, \ and\
  \bibinfo {author} {\bibfnamefont {S.}~\bibnamefont {Sachdev}},\ }\href
  {\doibase 10.1103/PhysRevB.78.085416} {\bibfield  {journal} {\bibinfo
  {journal} {Physical Review B}\ }\textbf {\bibinfo {volume} {78}},\ \bibinfo
  {pages} {085416} (\bibinfo {year} {2008})}\BibitemShut {NoStop}%
\bibitem [{\citenamefont {M\"{u}ller}\ \emph {et~al.}(2008)\citenamefont
  {M\"{u}ller}, \citenamefont {Fritz},\ and\ \citenamefont
  {Sachdev}}]{muller_quantum-critical_2008}%
  \BibitemOpen
  \bibfield  {author} {\bibinfo {author} {\bibfnamefont {M.}~\bibnamefont
  {M\"{u}ller}}, \bibinfo {author} {\bibfnamefont {L.}~\bibnamefont {Fritz}}, \
  and\ \bibinfo {author} {\bibfnamefont {S.}~\bibnamefont {Sachdev}},\ }\href
  {\doibase 10.1103/PhysRevB.78.115406} {\bibfield  {journal} {\bibinfo
  {journal} {Physical Review B}\ }\textbf {\bibinfo {volume} {78}},\ \bibinfo
  {pages} {115406} (\bibinfo {year} {2008})}\BibitemShut {NoStop}%
\bibitem [{\citenamefont {Principi}\ and\ \citenamefont
  {Vignale}(2015)}]{principi_violation_2015}%
  \BibitemOpen
  \bibfield  {author} {\bibinfo {author} {\bibfnamefont {A.}~\bibnamefont
  {Principi}}\ and\ \bibinfo {author} {\bibfnamefont {G.}~\bibnamefont
  {Vignale}},\ }\href {\doibase 10.1103/PhysRevLett.115.056603} {\bibfield
  {journal} {\bibinfo  {journal} {Physical Review Letters}\ }\textbf {\bibinfo
  {volume} {115}},\ \bibinfo {pages} {056603} (\bibinfo {year}
  {2015})}\BibitemShut {NoStop}%
\bibitem [{\citenamefont {Xie}\ and\ \citenamefont
  {Foster}(2016)}]{xie_transport_2016}%
  \BibitemOpen
  \bibfield  {author} {\bibinfo {author} {\bibfnamefont {H.-Y.}\ \bibnamefont
  {Xie}}\ and\ \bibinfo {author} {\bibfnamefont {M.~S.}\ \bibnamefont
  {Foster}},\ }\href {\doibase 10.1103/PhysRevB.93.195103} {\bibfield
  {journal} {\bibinfo  {journal} {Phys. Rev. B}\ }\textbf {\bibinfo {volume}
  {93}},\ \bibinfo {pages} {195103} (\bibinfo {year} {2016})}\BibitemShut
  {NoStop}%
\bibitem [{\citenamefont {Lucas}\ and\ \citenamefont
  {Fong}(2017)}]{lucas_hydrodynamics_2017}%
  \BibitemOpen
  \bibfield  {author} {\bibinfo {author} {\bibfnamefont {A.}~\bibnamefont
  {Lucas}}\ and\ \bibinfo {author} {\bibfnamefont {K.~C.}\ \bibnamefont
  {Fong}},\ }\href {http://arxiv.org/abs/1710.08425} {\bibfield  {journal}
  {\bibinfo  {journal} {arXiv:1710.08425 [cond-mat]}\ } (\bibinfo {year}
  {2017})}\BibitemShut {NoStop}%
\end{thebibliography}

%

\end{document}